# Topological acoustics

Zhaoju Yang[1], Fei Gao[1], Xihang Shi[1], Xiao Lin[1], Zhen Gao[1], Yidong Chong[1,2] and

Baile Zhang[1,2*]

[1]*Division of Physics and Applied Physics, School of Physical and Mathematical Sciences,*

*Nanyang Technological University, Singapore 637371, Singapore.*

[2]*Centre for Disruptive Photonic Technologies,*

*Nanyang Technological University, Singapore 637371, Singapore.*

*\*To whom correspondence should be addressed:*

*E-mail: blzhang@ntu.edu.sg*

**Abstract**

The manipulation of acoustic wave propagation in fluids has numerous applications, including in everyday life. Acoustic technologies have frequently developed in tandem with optics, using shared concepts such as waveguiding and meta-media. It is thus noteworthy that an entirely novel class of electromagnetic waves, known as 'topological edge states', has recently been demonstrated. These are inspired by the electronic edge states occurring in topological insulators, and possess a striking and technologically promising property: the ability to travel in a single direction along a surface without back-scattering, regardless of the existence of defects or disorder. Here we develop an analogous theory of topological fluid acoustics, and propose a scheme for realizing topological edge states in an acoustic structure containing circulating fluids. The phenomenon of disorder-free one-way sound propagation, which does not occur in ordinary acoustic devices, may have novel applications for acoustic isolators, modulators and transducers.

Since the 1980s, it has been known that the bands of certain insulators are 'topologically nontrivial': not smoothly deformable into the bands of a conventional insulator. Such systems, collectively referred to as 'topological insulators' [1-3], can exhibit edge states which propagate in a single direction along the edge of a two-dimensional sample. These states are 'topologically protected', meaning that they are tied to the topology of the underlying bands and cannot be eliminated by perturbations, and are hence immune to backscattering from disorder or shape variations. Some years ago, Haldane and Raghu [4] predicted that a similar phenomenon can arise in the context of classical electromagnetism [4-13], which was subsequently bourne out by experiments on microwave-scale magneto-optic photonic crystals [6] and other photonic devices[8,10,11].

In order to realize topological edge states using sound, we begin with a spatially periodic medium (in order to have band structures), and introduce a mechanism that breaks time-reveral symmetry (to allow for one-way propagation). A periodic acoustic medium, sometimes called a 'phononic crystal (PC)' [14], is commonly realized by engineering a structure whose acoustic properties (elastic moduli and/or mass density) vary periodically on a scale comparable to the acoustic wavelength. As for T-breaking, although traditional acoustic devices lack an efficient mechanism for accomplishing this, a recent breakthrough [15] has shown that strong T-breaking can be achieved in a 'meta-atom' containing a ring of circulating fluid. Although these developments have direct device applications as acoustic diode [16] and acoustic circulator [15], they do not have the topological protection against defects possessed by the topological edge states we will develop. We utilize the design concept by incorporating circulating fluid elements into a PC structure. As shall be seen, the resulting acoustic band structure is topologically nontrivial, and maps theoretically onto an integer quantum Hall gas [1]—the simplest version of a two-dimensional (2D) topological insulator.

We note also that several authors have studied topological vibrational modes in mechanical lattices [17-19]. The present system, by contrast, involves sound waves in continuous fluid media, which is considerably more relevant for existing acoustic technologies.

A schematic of the proposed system is shown in Fig. 1(a). It is a triangular lattice of lattice constant *a*, where each unit cell consists of a rigid solid cylinder (e.g., a metal cylinder) with radius $r_1$, surrounded by a cylindrical fluid-filled region of radius $r_2$. The remainder of the unit cell (i.e. the region of $r>r_2$) consists of a stationary fluid, separated from the fluid in the cylindrical region (i.e. the region of $r_1<r<r_2$) by a thin impedance-matched layer at radius $r_2$. (This layer can be achieved using a thin sheet of solid material that is permeable to sound.) The central cylinder rotates along its axis with angular speed $\Omega$, which produces a circulatory flow in the surrounding fluid in the region of $r_1<r<r_2$. (We will not consider the possibility of Taylor vortex formation [20] caused by large $\Omega$ in experiment because we here focus on 2D model and Taylor vortex does not contribute an effective flux through *x-y* plane.) We assume that fluid velocity is much slower than the speed of sound (Mach number of less than 0.3). The motion of the fluid can be described by a circulating 'Couette flow' distribution [20]; the velocity field points in the azimuthal direction, with component $v_\theta = -\frac{\Omega r_1^2}{r_2^2 - r_1^2} r + \frac{\Omega r_1^2 r_2^2}{r_2^2 - r_1^2} \frac{1}{r}$, where *r* is measured from the origin at the axis of the cylinder. This angular velocity is equal to $\Omega$ at radius $r=r_1$, and zero at radius $r=r_2$.

The propagation of sound waves in the presence of such a steady-state non-homogenous velocity background is described in Refs. [21, 22]. Assuming that the viscosity and heat flow are negligible, the waves obey a 'sound master equation'

$$\frac{1}{\rho}\nabla \cdot \rho \nabla \phi - (\partial_t + \vec{v}_0 \cdot \nabla)\frac{1}{c^2}(\partial_t + \vec{v}_0 \cdot \nabla)\phi = 0 \qquad (1)$$

where $\rho$ is the fluid density, $c$ is the speed of sound, and $\vec{v}_0$ is the background fluid velocity (i.e. the Couette flow distribution in the region of $r_1<r<r_2$ and stationary fluid in the region of $r>r_2$, where r is measured from the center of each unit cell). The relation between velocity potential $\phi$ and sound pressure $p$ is $p = \rho(\partial_t + \vec{v}_0 \cdot \nabla)\phi$. We model the surface of each cylinder as an impenetrable hard boundary by setting $\vec{n} \cdot \nabla \phi = 0$ where $\vec{n}$ is the surface normal vector. We restrict our attentions to time-harmonic solutions with frequency $\omega$ and neglect second order terms as $\left|\vec{v}_0/c\right|^2 \ll 1$. With a change of variables $\Psi = \sqrt{\rho}\phi$ the master equation can be rewritten as

$$[(\nabla - i\vec{A}_{eff})^2 + V(x,y)]\Psi = 0 \tag{2}$$

where the effective vector and scalar potentials are

$$\vec{A}_{eff} = -\frac{\omega \vec{v}_0(x,y)}{c^2} \tag{3}$$

$$V(x,y) = -\frac{1}{4}\left|\nabla \ln \rho\right|^2 - \frac{1}{2}\nabla^2 \ln \rho + \frac{\omega^2}{c^2} \tag{4}$$

Evidently Eq. (2) maps onto the Schrodinger equation for a spinless charged quantum particle in nonuniform vector and scalar potentials. For nonzero $\Omega$, the inner boundary of the Couette flow contributes positive effective magnetic flux, and the rest of the Couette flow contributes negative effective magnetic flux; the net magnetic flux, integrated over the entire unit cell, is zero. The acoustic system thus behaves like a 'zero field quantum Hall' system [23] and is periodic in the unit cell.

It is worth mentioning that a similar approach to construct an effective magnetic vector potential for classical wave propagation has been discussed by Berry and colleagues [24, 25]. These authors showed that an irrotational ('bathtub') fluid votex exhibits a classical wavefront dislocation effect, analogous to the Aharanov-Bohm effect. Here we advance this insight by applying the flow model to a PC context, so that the effective magnetic vector potential gives rise to a topologically nontrivial acoustic bandstructure.

From Eq. (1), we can calculate the acoustic bandstructures using the finite element method. For simplicity, we assume the fluids involved are air. The results, with $\Omega = 0$ and $\Omega \neq 0$, are shown in Fig. 1b (the lattice constant $a$ is set as 0.2 m). For $\Omega = 0$ [red curves in Fig. 1(b)], the acoustic bandstructure exhibits a pair of Dirac points at the corner of the hexagonal Brillouin zone, at frequency $\omega_0 = 0.577 \times 2\pi c_a / a$ (992 Hz), where $c_a$ is the sound velocity in air.

For $\Omega \neq 0$ the circulating air flow produces a dramatic change in the bandstructure [blue curves in Fig. 1(b)]. Here, we set the angular velocity of the inner rods to be $\Omega = 2\pi \times 400$ rad/s (achievable with miniature electric motors). The Dirac point degeneracies are lifted, producing a finite complete bandgap. The frequency splitting at the zone corners as a function of $\Omega$, is plotted in Fig. 1(c). The ratio of the operating frequency to the bandgap, which is an estimate for the penetration depth of the topological edge states in units of the lattice constant, is on the order of $\omega/\delta\omega \approx 10$ for the range of angular velocities plotted here. For $\Omega = 2\pi \times 400$ rad/s, the band gap ranges from 914 Hz to 1029 Hz, corresponding to a relatively narrow bandwidth of 115 Hz.

Each acoustic band can be characterized by a topological invariant, the Chern number [4]. The Berry connection and Chern number of the $n$th acoustic band can be defined as follows:

$$\vec{\mathcal{A}}_n = i \langle \phi_{nk} | \nabla_{\vec{k}} | \phi_{nk} \rangle \tag{5}$$

$$C_n = \frac{1}{2\pi} \iint_{BZ} (dk_a \wedge dk_b) \nabla_{\vec{k}} \times \vec{\mathcal{A}}_n \tag{6}$$

We have numerically verified that the two bands in Fig. 1(b), split by the T-breaking, have Chern numbers of $\pm 1$. The principle of 'bulk-edge correspondence' then predicts that, for a finite PC the gap between these two bands is spanned by unidirectional acoustic edge states, analogous to the electronic edge states occurring in the quantum Hall effect [26].

To confirm the existence of these topologically-protected acoustic edge states, we numerically calculate the bandstructure for a 20×1 super-cell [27] (a ribbon that is 20-unit-cell wide in *y* direction and infinite along *x* direction. As shown in Fig. 2(a), for $\Omega = 2\pi \times 400$ rad/s the bandgap contains two sets of edge states, which are confined to opposite edges of the ribbon and have opposite group velocities.

Figs. 2(b-c) show the propagation of these edge states in a finite (34×14) lattice. In these simulations, the upper edge of the PC is enclosed by a sound-impermeable hard boundary (e.g., a flat metal surface), in order to prevent sound waves from leaking into the upper half space; absorbing boundary conditions are applied to the sides. A point sound source with mid-gap frequency $\omega_0$ is placed near the upper boundary. For $\Omega = 2\pi \times 400$ rad/s, this excites a unidirectional edge state which propagates to the left along the interface [Fig. 2(b)]. If the sign of angular velocity is reversed, the edge state would be directed to the right (not plotted). The field distribution for a reduced angular velocity $\Omega = 2\pi \times 200$ rad/s [Fig. 2(c)] shows an edge state with a longer penetration depth because of a narrower bandgap.

Due to the lack of backward-propagating edge modes, the presence of disorder cannot cause backscattering. Fig. 3(a) shows an acoustic cavity located along the interface; the incident wave flows through the cavity, and excites localized resonances within the cavity, but does not backscatter. Fig. 3(b) shows a Z-shape bend connecting two parallel surfaces at different *y*; again, the acoustic edge states are fully transmitted across the bend. Finally, Fig. 3(c) shows a 180-degree bend which allows acoustic edge states to be guided from the top of a sample to the bottom of the sample; note that the left boundary in this sample is a zig-zag boundary, which supports one-way edge states with different dispersion relations.

Our proposed system should be quite practical to realize. Similar effects can be achieved with alternative designs featuring circulatory fluid velocity distributions; e.g., having azimuthally-directed fans in each unit cell [15], or stirring with a rotating disc on the top plate [28]. The effect could be largely tunable from audible to even ultrasonic frequencies by appropriately scaling down lattice constant or practically operating at higher band gaps with larger Chern number. Acoustic devices based on these topological properties can be useful for invisibility from sonar detection, one-way signal processing regardless of disorders, environmental noise control, which will greatly broaden our interest in military, medical and industrial applications.

This work was sponsored by Nanyang Technological University under Start-Up Grants, and Singapore Ministry of Education under Grant No. Tier 1 RG27/12 and Grant No. MOE2011-T3-1-005. CYD acknowledges support from the Singapore National Research Foundation under grant No. NRFF2012-02.

# Figure legends

FIG. 1 (color online). A two-dimensional acoustic topological insulator and its band structure. (a) Triangular acoustic lattice with lattice constant $a$. $a=0.2$ m in the following calculation. Inset: unit cell containing a central metal rod of radius $r_1 = 0.2a$, surrounded by an anticlockwise circulating fluid flow (flow direction indicated by red arrows) in a cylinder region of radius $r_2 = 0.4a$. (b) Band structures of the acoustic lattice without the circulating fluid flow (red curves; $\Omega = 2\pi \times 0$ rad/s) and with fluid flow (blue curves; $\Omega = 2\pi \times 400$ rad/s). In the gapped band structure, the bands have Chern number $\pm 1$ (blue labels). Left inset: enlarged view of Dirac cone. Right lower inset: the first Brillouin zone. (c) Frequency splitting as a function of the angular velocity of the cylinder in each unit cell. The degeneracy at the Dirac point with frequency $\omega_0 = 0.577 \times 2\pi c_a / a$ (992 Hz) is removed for $\Omega \neq 0$.

FIG. 2 (color online). Acoustic one-way edge states. (a) Dispersion of the one-way acoustic edge states (red curves) occurring in a finite strip of the acoustic lattice, for $\Omega = 2\pi \times 400$ rad/s. The left and right red curves correspond to edge states localized at the bottom and top of the strip. (b-c) The normalized acoustic pressure $p$ for a left-propagating acoustic edge state at frequency $\omega_0 = 0.577 \times 2\pi c_a / a$ (992 Hz) for $\Omega = 2\pi \times 400$ rad/ (b) and $\Omega = 2\pi \times 200$ rad/s (c). Lattice parameters are the same as in FIG. 1.

FIG. 3 (color online). Demonstration of the robustness of acoustic one-way edge states against disorder. Topological protection requires the waves to be fully transmitted through an acoustic cavity (a), a Z-shape bend along the interface (b) and a 180-degree bend (c). The operating frequency is $\omega_0 = 0.577 \times 2\pi c_a / a$ (992 Hz) and $\Omega = 2\pi \times 400$ rad/s. Lattice parameters are the same as in FIG. 1.

Figure 1

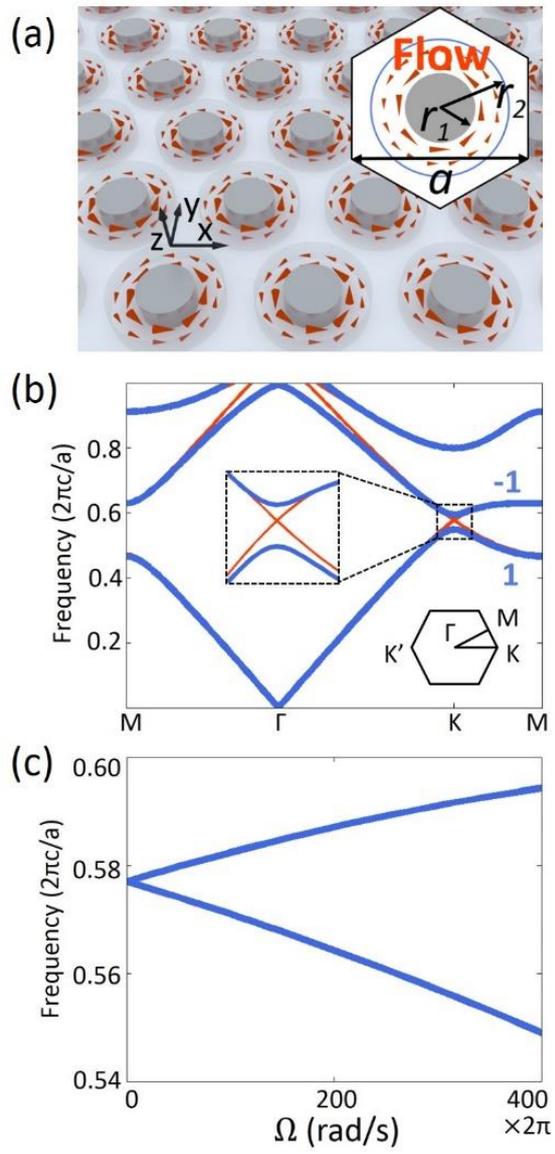

Figure 2

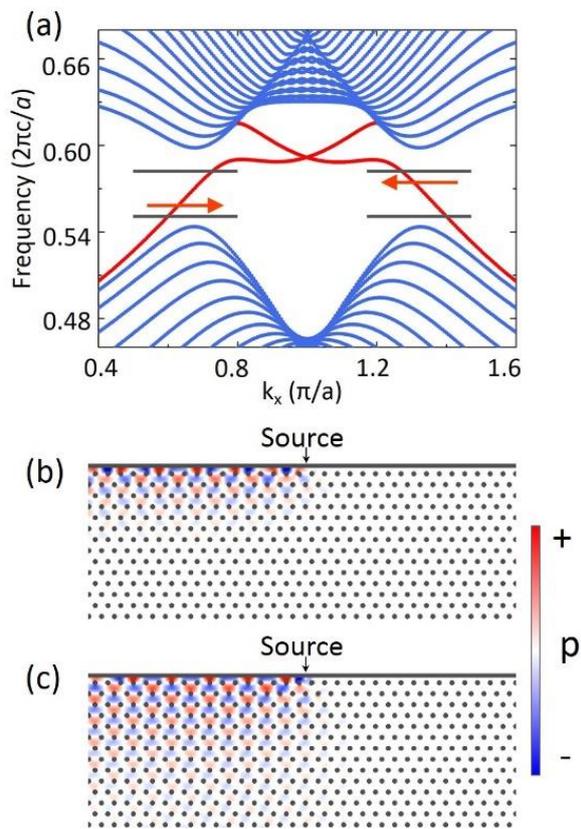

Figure 3

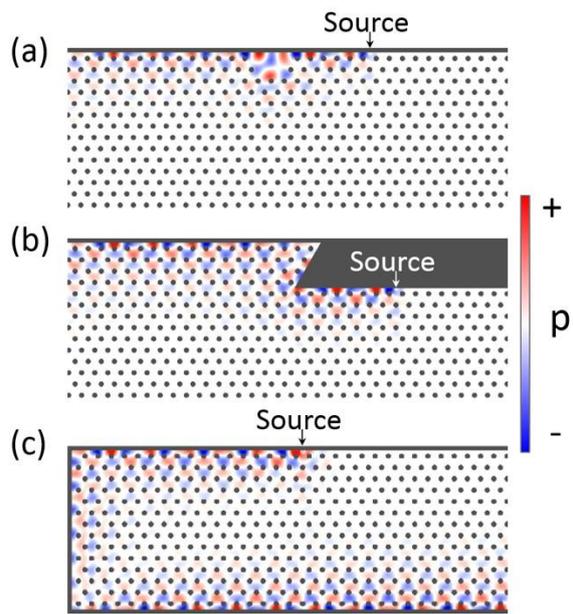